\documentclass[conference,10pt]{IEEEtran}
\usepackage{verbatim}  
\usepackage{epsf}
\usepackage{epsfig}
\usepackage{epstopdf}
\usepackage{enumerate}
\usepackage{times}
\usepackage{amsmath}
\usepackage{amssymb}
\usepackage{cite}

\IEEEoverridecommandlockouts

\usepackage{algorithm}
\usepackage{algpseudocode}

\usepackage[hidelinks,bookmarks=false]{hyperref}

\title{Total Power Minimization: Joint Antenna Selection and Beamforming Design}

\author{\IEEEauthorblockN{Mostafa Medra$^\ast$ \quad Andrew W. Eckford$^\dagger$ \quad Raviraj Adve$^\ast$}
\IEEEauthorblockA{$^\ast$ Department of Electrical and Computer Engineering, University of Toronto, ON, Canada. \\
$^\dagger$ Department of Electrical Engineering and Computer Science, York University, Toronto, ON, Canada.}
\thanks{This research was supported by TELUS Canada and the Natural Sciences and Engineering Research Council (NSERC) of Canada.}
}

\begin{document}
\maketitle

\begin{abstract}

In this paper, we consider the total  power minimization problem when we have signal-to-interference-plus-noise ratio (SINR) constraints. The consumed power in the circuits depends on the number of active antennas, which can be modeled using zero-norm. Due to the difficulty of dealing with the non-convex zero-norm, we used the standard alternate weighted one-norm approach. We addressed the total power minimization for a narrowband system with and without per-antenna power constraints (PAPCs).  We derived iterative closed-form expressions in both cases.  Then we analysed the case when we have multiple bands operating at the same time. Analogous closed-form expressions are provided.  Our simulation results show that significant gains can be obtained in terms of the total power required compared to standard methods that do not take into account the circuit power.

\end{abstract}

\section{Introduction}

When several antennas are available at a base station (BS), those antennas allow the BS to serve multiple users simultaneously; e.g., \cite{ShiftingtheMIMO,Reference2,Jointoptimal,Linearprecoding,SymbollevelandMulticast,Iterativemultiuseruplink,OptimalMultiuserTransmit}. The BS serves multiple users through the use of beamforming techniques. Since the computational complexity is an important factor in design, typically linear beamforming techniques are used.  Among conventional linear precoders are the maximum ratio transmission (MRT) \cite{Maximumratiotransmission}, zero-forcing (ZF) \cite{Zeroforcingmethods} and regularized ZF (RZF) \cite{Avectorperturbation}.  Those beamformers are of low-computational complexity and can be obtained  in closed-form expressions.  When the users have single antenna each, the beamformers that minimize the transmission power subject to certain signal-to-interference-plus-noise ratio (SINR) constraints can be optimally found \cite{Reference2,OptimalMultiuserTransmit}. That problem can be formulated as a convex problem that can be efficiently solved. The KKT conditions of which also allow for an iterative closed-form expressions \cite{OptimalMultiuserTransmit}.

In practice, each antenna at the BS will be driven by its own power amplifier, and, hence, constraints on the  transmitted power from each antenna should be included to the beamforming design. When per-antenna power constraints (PAPCs) are introduced to the power minimization problem under SINR constraints, the problem remains convex since the PAPCs are convex. Using the KKT conditions, iterative closed-form solutions can be obtained \cite{TransmitterOptimization,LowComplexityRobustMISOPAPCs}. The PAPCs can be directly applied when the system is a norrowband system. Systems operating simultaneously on many bands; e.g., OFDM systems, would have an IFFT operation changing the power distribution across the antennas.

The problem of power minimization under certain SINR constraints is well-studied. However, most of the formulations of this problem ignore the power consumed in the RF circuits, and focus only on the transmitted power. Realistic power consumption models that also
take the hardware-consumed power into account were studied; e.g.,  \cite{MassiveMIMOforenergy,Modelingthehardwarepower}. While the transmit power required to meet fixed SINR constraints decays in an inverse fashion to the number of BS antennas~\cite{EnergyandSpectralEfficiency,MassiveMIMOintheUL}, the RF circuit power is linear in this number. There is, therefore, an optimal number of antennas that requires the least possible total power~ \cite{MassiveMIMOforenergy,Modelingthehardwarepower}.  In~\cite{Energyefficiencyanalysis}, the authors showed that the energy efficiency is is a quasi-concave function of the number of BS antennas in the case of  massive MIMO downlink system.

Introducing circuit power minimization into the power minimization problem is challenging because the consumed power depends on the number of active antennas. While the sparsity of a signal can be determined by using the zero-norm ($\ell_{0}$ norm) of the power on each antenna, such a norm is not convex and the associated problems are NP hard; e.g.,~\cite{Decodingblinearprogramming, Robustuncertaintyprinciples}. One possible approach is to replace the zero-norm with a weighted one-norm. We will show that we can transform the total power minimization problem using weighted one-norm  into a convex iterative problem that can be efficiently solved. The problem can be viewed as jointly solving the antenna selection (e.g., \cite{MIMOsystemswithantenna,AntennaSelectionStrategiesfor}) and beamforming design, when the objective is power minimization. This problem was solved for fixed beamforming directions (ZF and MRT) when the number of antennas is large~\cite{JointTransmitandCircuitPower}. The analysis therein is based on asymptotic results and only provides the number of antennas, whereas our work is applicable to any number of available antennas and is capable of determining which antennas are on. The $\ell_{0}$ norm was used for minimizing the number of antennas for a given rate constraints in \cite{Minimalantennasubsetselection}. In~\cite{ImpactofTransceiverPower}, for a given uplink sum rate, the authors investigated the optimal numbers of BS antennas and users for a single-cell system.  The number of active antennas was also considered in a point-to-point large-scale MIMO channel in ~\cite{HowmanyRFchains}. The minimum BS power consumption for a given sum rate in a large scale MIMO system is addressed in ~\cite{Minimumpowerconsumptionofabase}.

In this paper, we will first review the quality-of-service (QoS) problem for a given set of  SINR targets and the closed-form expressions for that problem. We will also review the required modifications for the PAPCs case. Then, we will show how the circuit power minimization can be introduced using the weighted one-norm approach. We will formulate the problem as a convex problem, and we will show that by analysing the KKT, we can obtain closed-form expressions for the optimal beamformers.  We will then introduce the PAPCs, and show how to modify the closed-form expressions to accommodate for the extra constraints. For the multi-band case, we will show that convex formulations that allow for closed-form expressions are still possible. In the simulation section, we will show the trade-off between the number of antennas and the total consumed power.  We show that significant gains can be obtains by turning many antennas off for the single narrow band case even when PAPCs are on. We also show that gains can be obtained even when many narrow bands are on.

\section{System Model}

We consider a  multiple-input single-output (MISO) downlink system where $K$ users, each with a single antenna, are served by a base station (BS) with $N_t$ antennas. We assume that the BS is provided with perfect channel state information of the users. We let $\mathbf{h}_k \in \mathbb{C}^{N_t}$  denote the channel between the BS and user $k$, and let $s_k$ denote the intended normalized data symbol for that user. We design the precoding vector $\mathbf{w}_k$ for transmission from the BS to user $k$. When the BS transmits $\sum_{k=1}^K\mathbf{w}_k s_k$, we can write the received signal at user $k$ as:
\begin{equation}\
    y_{k}= \mathbf{h}_{k}^H \mathbf{w}_{k} s_{k} + \textstyle\sum_{i \neq k}\mathbf{h}_{k}^H \mathbf{w}_{i} s_{i} + n_{k},
\end{equation}
in which $n_{k} $ is zero-mean circular Gaussian noise of variance $\sigma_{k}^2$. We will express each user's QoS constraint by an SINR constraint: $\mathsf{SINR}_k\geq \gamma_k$.
This SINR constraint can be written as
\begin{equation}\
    \mathsf{SINR}_{k}= \frac{\mathbf{h}_{k}^H \mathbf{w}_{k} \mathbf{w}_{k}^H \mathbf{h}_{k}}{\mathbf{h}_{k}^H ( \textstyle\sum_{i \neq k}\mathbf{w}_{i} \mathbf{w}_{i}^H) \mathbf{h}_{k}+ \sigma_{k}^2}\geq \gamma_k,
\end{equation}
or equivalently $\mathbf{h}_{k}^H \mathbf{Q}_{k} \mathbf{h}_{k} - \sigma_k^2 \geq 0$, where
\begin{equation}
\mathbf{Q}_{k}=\mathbf{w}_k \mathbf{w}_k^H/\gamma_k-\sum_{i \neq k}\mathbf{w}_i \mathbf{w}_i^H.
\end{equation}

When we deal with many narrow bands, we will use the superscript $j$ to indicate the narrow band index, and $N_b$ as the number of narrow bands. For single bands systems, the index $j$ will be dropped to simplify the notations.

If we denote the signal transmitted from antenna $i$ by $x_i$, then the average transmitted  power from the BS can be written as $\textstyle\sum_{i=1}^{N_t} E |x_i|^2 $. In the case of zero-mean independent data symbols of normalized power, this  becomes $\textstyle\sum_{k=1}^K  \mathbf{w}_k^H \mathbf{w}_k$, and the average transmitted power from the $i$th antenna is $\Bigl[\textstyle\sum_{k=1}^K \mathbf{w}_k \mathbf{w}_k^H \Bigr]_{i,i}$. The BS has to minimize  the consumed power while satisfying the SINR constraints.

The conventional problem of minimizing the transmitted beamforming  power under SINR constraints can, therefore, be written as
\begin{subequations}\label{perfect_csi}
\begin{align}
     \min_{\substack{\mathbf{w}_k}} \quad & \sum_k \mathbf{w}_k^H \mathbf{w}_k  \\
    \text{s.t.} \quad & \mathbf{h}_{k}^H \mathbf{Q}_{k} \mathbf{h}_{k} - \sigma_k^2 \geq 0, \quad \forall k.
   \end{align}
\end{subequations}
This formulation is not convex. However, since the SINR expression does not change with the phase of $\mathbf{w}_{k}$, we can express the SINR constraint as
$$\mathbf{w}_k^H  \mathbf{h}_k \sqrt{1+1/\gamma_k} \geq  \sqrt{\sum_i |\mathbf{w}_i^H  \mathbf{h}_k|^2 + \sigma_k^2 }.$$
This equivalent formulation is a convex conic constraint that can be efficiently solved; e.g., using CVX tool~\cite{cvx} accessible from $\textsc{MATLAB}$. The KKT conditions of this problem allow for closed-form expressions for the optimal beamformers as well \cite{Reference2,Jointoptimal}.

When PAPCs are introduced, the power minimization problem can be written as
\begin{subequations}\label{PAPCs}
\begin{align}
     \min_{\substack{\mathbf{w}_k}} \quad &  \sum_k \mathbf{w}_k^H \mathbf{w}_k  \\
    \text{s.t.} \quad &  \mathbf{h}_{k}^H \mathbf{Q}_{k} \mathbf{h}_{k} - \sigma_k^2 \geq 0, \quad \forall k. \\
    \quad &  p_a \geq \left[ \sum_k \mathbf{w}_k \mathbf{w}_k^H\right]_{i,i},  \quad \forall i,
   \end{align}
\end{subequations}
where $p_a$ is the PAPC. Closed-form iterative solutions are available as well~\cite{TransmitterOptimization,LowComplexityRobustMISOPAPCs}.

\section{Narrowband Total Power Minimization}

As we can see, the problem in \eqref{perfect_csi} does not take into account the power consumed in the circuits driving the antennas. Such a number depends on how many antennas are active.  Accordingly,  the number of active antennas is then $\sum_i \| P_i\|_0$, where $P_i$ is the power emitted from the $i$th antenna. If we assume that the the power consumed per antenna is denoted by $c_1$ and $c_2$ models the amplifier efficiency, then the more general problem of minimizing the \emph{total consumed power} while satisfying the SINR constraints can be written as
\begin{subequations}\label{total_power_min}
\begin{align}
     \min_{\substack{\mathbf{w}_k}, P_i} \quad & c_1 \sum_i \| P_i\|_0 +c_2 \sum_k \mathbf{w}_k^H \mathbf{w}_k  \\
    \text{s.t.} \quad &  \mathbf{h}_{k}^H \mathbf{Q}_{k} \mathbf{h}_{k} - \sigma_k^2 \geq 0, \quad \forall k. \\
    \quad &  P_i \geq \left[ \sum_k \mathbf{w}_k \mathbf{w}_k^H\right]_{i,i},  \quad \forall i.
   \end{align}
\end{subequations}
Since $\ell_0$ norm is not convex and hard to deal with, we will use the standard weighted one-norm approach. Using that approach, \eqref{total_power_min} can be written as
\begin{subequations}\label{one_norm_pmin}
\begin{align}
     \min_{\substack{\mathbf{w}_k}, P_i} \quad & c_1 \sum_i s_i P_i +c_2 \sum_k \mathbf{w}_k^H \mathbf{w}_k  \\
    \text{s.t.} \quad &  \mathbf{h}_{k}^H \mathbf{Q}_{k} \mathbf{h}_{k} - \sigma_k^2 \geq 0, \quad \forall k. \label{total_power_min_c1} \\
    \quad &  P_i \geq \left[ \sum_k \mathbf{w}_k \mathbf{w}_k^H\right]_{i,i},  \quad \forall i. \label{total_power_min_c2}
   \end{align}
\end{subequations}
The weighted one-norm updates the weights $s_i$ iteratively as $s_i=1/(P_i+ \delta),$ where $\delta$ is a regularization constant.

If we let $\nu_k$ and $\lambda_i$ denote the dual variable of the $k$th constraint in \eqref{total_power_min_c1}, and the $i$th constraint in \eqref{total_power_min_c2}, respectively, then we can write the Lagrangian of \eqref{one_norm_pmin}  as
\begin{multline}
     \mathcal{L}(\mathbf{w}_k, P_i, \nu_k, \lambda_i)=c_1 \sum_i s_i P_i +c_2 \sum_k \mathbf{w}_k^H \mathbf{w}_k \\ - \sum_k \nu_k ( \mathbf{h}_{k}^H \mathbf{Q}_{k} \mathbf{h}_{k} - \sigma_k^2)- \sum_i \lambda_i \Bigl(P_i - \left[ \sum_k \mathbf{w}_k \mathbf{w}_k^H\right]_{i,i}\Bigr). \nonumber
\end{multline}
From the KKT conditions, we have that $c_1  s_i = \lambda_i$. If we let $\Lambda$ denote a diagonal matrix whose $(i,i)$th element is $\lambda_i$ then the beamforming vector should satisfy the KKT condition
\begin{equation}\label{w_eqn}
c_2 \mathbf{w}_k =\Biggl( \frac{\nu_k}{\gamma_k}\mathbf{h}_{k} \mathbf{h}_{k}^H-\sum_{j\neq k} \nu_j \mathbf{h}_{j} \mathbf{h}_{j}^H  - \Lambda \Biggr)\mathbf{w}_k,
\end{equation}
which can be rearranged, similar to \cite{Reference2},  to show that the dual variables $\{\nu_k\}$ should satisfy the fixed-point equations
\begin{equation}\label{nu}
\nu_k^{-1}  = \mathbf{h}_{k}^H \Bigl(c_2\mathbf{I}+\textstyle\sum_{j} \nu_j \mathbf{h}_{j} \mathbf{h}_{j}^H +  \Lambda \Bigr)^{-1} \mathbf{h}_{k}  \Bigl(1+\frac{1}{\gamma_k} \Bigr).
\end{equation}
Once we obtain $\nu_k$, we can solve \eqref{w_eqn} as an eigen equation to obtain the directions. The power loading; $\|\mathbf{w}_k\|$, can be obtained from solving the $K$ SINR constraints satisfied by equality at optimality. If this is not the case for constraint $k$, then we can decrease $\mathbf{w}_k$, which will still satisfy all the constraints, and provide a lower objective, which contradicts the presumed optimality. These steps are summarized in Alg.~\ref{Alg1}. Detailed complexity analysis and implementation issues are addressed for a similarly structured problem in~\cite{OffsetBasedBeamforming}.

 \begin{algorithm}
\caption{Narrowband total power minimization}
\label{Alg1}
\begin{algorithmic}[1]
\State Initialize $s_k=1$ and obtain the corresponding $\Lambda$.
 \For{\texttt{A certain number of iterations}}
\State  Solve the fixed-point equations in \eqref{nu} to obtain $\nu_k$.
  \State Solve the eigen equation in \eqref{w_eqn} to find the direction of $\mathbf{w}_k$.
  \State Solve the $K$ linear equations arrising from \eqref{total_power_min_c1} holding with equality at optimality to obtain the power loading; $\|\mathbf{w}_k\|$.
  \State Update the weights $s_i=1/(P_i+ \delta)$ and $\Lambda$.
 \EndFor
\end{algorithmic}
\end{algorithm}

\section{Narrowband Total Power Minimization under per-Antenna Power Constraints}

The previous section focused on the case of minimizing  the total power with no individual antenna power constraints. However, in practice, each antenna is driven by its own power amplifier, and the addition of PAPCs is inevitable.
In this section, we will show how the previous analysis and algorithm can be tailored to address the  PAPCs case as well.

The problem in \eqref{one_norm_pmin} can be modified to include PAPCs as follows
\begin{subequations}\label{one_norm_papc}
\begin{align}
     \min_{\substack{\mathbf{w}_k}, P_i} \quad & c_1 \sum_i s_i P_i +c_2 \sum_k \mathbf{w}_k^H \mathbf{w}_k  \\
    \text{s.t.} \quad &  \mathbf{h}_{k}^H \mathbf{Q}_{k} \mathbf{h}_{k} - \sigma_k^2 \geq 0, \quad \forall k.  \label{one_norm_papc_c1}\\
    \quad &  P_i \geq \left[ \sum_k \mathbf{w}_k \mathbf{w}_k^H\right]_{i,i},  \quad \forall i. \\
     \quad &  p_a \geq \left[ \sum_k \mathbf{w}_k \mathbf{w}_k^H\right]_{i,i},  \quad \forall i.
   \end{align}
\end{subequations}
Following a similar KKT analysis, if we let $q_i$ denote the dual variable for the PAPC, $\hat{\mathbf{Q}}$ denote a diagonal matrix whose $(i,i)$th element is $q_i$, then the Lagrangian of the problem in \eqref{one_norm_papc} can be written as
\begin{multline}
 \mathcal{L}(\mathbf{w}_k, P_i, \nu_k, \lambda_i, q_i)=c_1 \sum_i s_i P_i +c_2 \sum_k \mathbf{w}_k^H \mathbf{w}_k \\ - \sum_k \nu_k ( \mathbf{h}_{k}^H \mathbf{Q}_{k} \mathbf{h}_{k} - \sigma_k^2)- \sum_i \lambda_i \Bigl(P_i - \left[ \sum_k \mathbf{w}_k \mathbf{w}_k^H\right]_{i,i}\Bigr) \\
 - \sum_i q_i \Bigl(p_a - \left[ \sum_k \mathbf{w}_k \mathbf{w}_k^H\right]_{i,i}\Bigr).
\end{multline}

Similar to the previous case, we can observe from the KKT conditions that $c_1  s_i = \lambda_i$. Also the beamforming vector should satisfy the KKT condition
\begin{equation}\label{w_eqn2}
c_2 \mathbf{w}_k =\Biggl( \frac{\nu_k}{\gamma_k}\mathbf{h}_{k} \mathbf{h}_{k}^H-\sum_{j\neq k} \nu_j \mathbf{h}_{j} \mathbf{h}_{j}^H  - \Lambda -\hat{\mathbf{Q}} \Biggr)\mathbf{w}_k,
\end{equation}
which can be similarly rearranged to show that the dual variables $\{\nu_k\}$, in the PAPC case, should satisfy the fixed-point equations
\begin{equation}\label{nu2}
\nu_k^{-1}  = \mathbf{h}_{k}^H \Bigl(c_2\mathbf{I}+\textstyle\sum_{j} \nu_j \mathbf{h}_{j} \mathbf{h}_{j}^H +  \Lambda + \hat{\mathbf{Q}} \Bigr)^{-1} \mathbf{h}_{k}  \Bigl(1+\frac{1}{\gamma_k} \Bigr).
\end{equation}
And once the beamforming directions are obtained, the power loading can be calculated assuming that the SINR constraints hold with equality at optimality.

The above derivations are based on the values of the dual variable $q_i$, which are not known in advance. However, we can use efficient subgradient algorithms to obtain these values in an iterative way.  Subgradient algorithms and their convergence analysis were proposed in \cite{TransmitterOptimization}, and parameter selection and convergence speed were enhanced in \cite{LowComplexityRobustMISOPAPCs}. The basic idea of such algorithm is to solve the KKT conditions for an initial value of $q_i$, then check the PAPCs. If they are met, the algorithm terminates, other wise, the values of $q_i$ are updated. We update $q_i$ (or $\hat{\mathbf{Q}}$) such that  \cite{LowComplexityRobustMISOPAPCs}
\begin{equation}\label{qupdate_gen}
 q_i^{n+1}=\max\Bigl( q_i^{n}+ t_n  \Bigl(\left[ \sum_k \mathbf{w}_k \mathbf{w}_k^H\right]_{i,i}- p_a \Bigr), 0\Bigr),
\end{equation}
where  $t_n$ is the step size, and $n$ is the iteration index. The maximum operator guarantees that when the PAPC is not active, then the corresponding $q_i$ is zero. We can summarize the proposed algorithm as shown in Alg.~\ref{Alg2}. The complexity analysis in each iteration is the same as that of Alg.~\ref{Alg1}.

 \begin{algorithm}
\caption{Power minimization under PAPCs}
\label{Alg2}
\begin{algorithmic}[1]
\State Initialize $q_i=0$, $s_k=1$ and obtain the corresponding $\Lambda$.
 \For{\texttt{A certain number of iterations}}
\While { $ p_a < \left[ \sum_k \mathbf{w}_k \mathbf{w}_k^H\right]_{i,i}$ for any $i$}
\State  Solve \eqref{nu2} to obtain $\nu_k$.
  \State Solve  \eqref{w_eqn2} to find the direction of $\mathbf{w}_k$.
  \State Solve the $K$ linear equations arrising from \eqref{one_norm_papc_c1} holding with equality at optimality to obtain the power loading; $\|\mathbf{w}_k\|$.
  \State  Update $\hat{\mathbf{Q}}^{n+1}$ using \eqref{qupdate_gen}.
  \EndWhile
  \State Update the weights $s_i=1/(P_i+ \delta)$ and $\Lambda$.
\EndFor
\end{algorithmic}
\end{algorithm}


\section{Multi-Band Case}

The previous sections dealt with the single-band case. However, to turn an antenna off in the multi-band case, that antenna should have no power over all the beamformers of the different bands. Accordingly, if we let $P_i$ denote the power over all the $N_b$ narrow bands, instead of only one band, then we can use the weighted one-norm approach to solving \eqref{total_power_min}  as follows
\begin{subequations}
\begin{align}
     \min_{\substack{\mathbf{w}_k^j}, P_i} \quad & c_1 \sum_i s_i P_i +c_2 \sum_{k,j} (\mathbf{w}_k^j)^H \mathbf{w}_k^j  \\
    \text{s.t.} \quad &  (\mathbf{h}_k^j)^H \mathbf{Q}_k^j \mathbf{h}_{k}^j - (\sigma_k^j)^2 \geq 0, \quad \forall k, j.  \\
    \quad &  P_i \geq \left[ \sum_{k,j} \mathbf{w}_k^j (\mathbf{w}_k^j)^H\right]_{i,i},  \quad \forall i.
   \end{align}
\end{subequations}

Similar to the narrowband case, from the KKT conditions, we have that $c_1  s_i = \lambda_i$ and the beamforming vector satisfies
\begin{equation}\label{w_eqn_m}
c_2 \mathbf{w}_k^j =\Biggl( \frac{\nu_k^j}{\gamma_k^j}\mathbf{h}_{k}^j (\mathbf{h}_{k}^j)^H-\sum_{i\neq k} \nu_i^j \mathbf{h}_i^j (\mathbf{h}_i^j)^H  - \Lambda \Biggr)\mathbf{w}_k^j,
\end{equation}
where the dual variables $\{\nu_k^j\}$ satisfy
\begin{equation}\label{nu_m}
(\nu_k^j)^{-1}  = (\mathbf{h}_{k}^j)^H \Bigl(c_2\mathbf{I}+\textstyle\sum_{i} \nu_i^j \mathbf{h}_i^j (\mathbf{h}_i^j)^H +  \Lambda \Bigr)^{-1} \mathbf{h}_{k}^j  \Bigl(1+\frac{1}{\gamma_k^j} \Bigr). \nonumber
\end{equation}
The power loading for the beamformers of the $j$th narrowband is then obtained from solving the $K$ SINR constraints of that band satisfied by equality at optimality.

\section{Simulation Results}

In this section, we illustrate the performance of the proposed approaches in solving the total power minimization problem. We consider a system consisting of a BS with $N_t$ antennas, serving $K=4$ users per band. Fading is modelled using the standard Rayleigh model. We assume an SINR target of $\gamma=3$dB for all users and  that each user has normalized noise power; $\sigma_k^2=1$.  We set $c_1=0.3$ W/antenna and the amplifier efficiency to $30\%$; i.e., $c_2=1/0.3$. For the PAPCs, $P_a=0.4$. For the weighted one-norm, we used a regularization factor of $\delta=10^{-4}$, and 6 iterations.

\begin{figure}
\begin{center}
    \epsfysize= 2.0in
     \epsffile{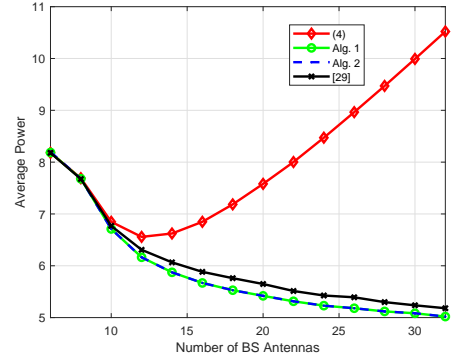}
\caption{The average total power  versus the number of BS antennas for $K=4$ users, $\gamma$=3dB, and $\sigma^2$=1.}\label{fig1}
\vspace{-.8cm}
\end{center}
\end{figure}

In Fig.~\ref{fig1}, we plot the total power versus the number of BS antennas. As we can see for \eqref{perfect_csi}, the reduction in the transmitted power is significant when the number of antennas is low, then the circuit power dominates  which causes the total power to grow almost linearly. We observe that  Alg.~\ref{Alg1} and Alg.~\ref{Alg2} are not able to reduce the power by turning off antennas when the number of antennas is low, however, as the number of antennas increases, they provide significant gains compared to the naive approach in \eqref{perfect_csi}. As a benchmark, we also compare with the algorithm that uses antenna selection as descriped in \cite{Fastalgorithmforantennaselection}, then solves the beamforming problem using \eqref{perfect_csi}. The antenna selection in \cite{Fastalgorithmforantennaselection} is based on finding the most correlated rows of the channel matrix then deleting the row of lower power. We keep deleting rows and calculating the power required using \eqref{perfect_csi} till the minimum power is obtained. Since this method deletes one row at a time, it takes more iterations than our proposed methods that work with a fixed number of iterations. In addition, our methods provide lower average power and is easily extendable to the case of PAPCs or multi-band. In table~\ref{table1}, we list the average number of active antennas versus the total number of antennas. Note that our closed-form expressions provide the same results obtained by solving \eqref{one_norm_pmin} or \eqref{one_norm_papc} using CVX tool~\cite{cvx}. We note that the problem with PAPCs can be infeasible. Using higher number of antennas can lower the transmitted power to avoid infeasibility.

\begin {table}
\begin{center}
\caption{Average number of active antennas}
\begin{tabular}{|c|c|c|c|c|c|c|c|}
  \hline
   $N_t$ & 8 & 12 & 16 & 20 & 24 & 28 & 32 \\ \hline
  Alg.~\ref{Alg1} &  7.41 &   8.82  &  9.22   & 9.32  &  9.33  &  9.27  &  9.30 \\ \hline
  Alg.~\ref{Alg2} & 7.47 &   8.83  &  9.22  &  9.32 &   9.33  &  9.27 &   9.30  \\ \hline
  [29] &  7.80  &  9.91   & 9.87  &  9.71  &  9.36  &  9.12  &  9.01  \\
  \hline
\end{tabular}
\label{table1}
\end{center}
\end {table}

\begin{figure}
\begin{center}
    \epsfysize= 2.0in
     \epsffile{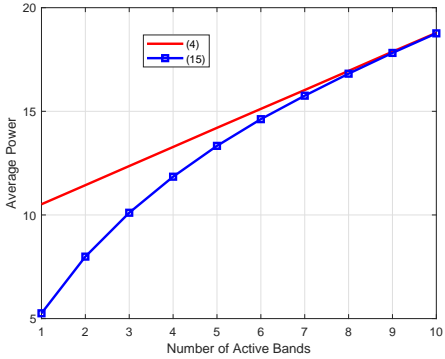}
\caption{The average power versus the  number of active bands for $K=4$ users per band, and an $N_t=32$ antennas BS.}\label{fig2}
\vspace{-.6cm}
\end{center}
\end{figure}

In Fig.~\ref{fig2}, we plot the total power  versus the number of operating bands for $K=4$ users per band, and an $N_t=32$ antennas. We can see that up to a few bands, the algorithm is still able to obtain gains by turning antennas off. In table~\ref{table2}, we list the average number of active antennas versus the total number of active bands. While the power gains are smaller for higher number of active bands, the smaller number of active antennas means that the computations would be easier and less power consuming.

\begin {table}
\begin{center}
\caption{Average number of active antennas}
\begin{tabular}{|c|c|c|c|c|c|c|c|c|c|c|}
   \hline
   $N_b$ & 1 & 2 & 3 & 4 & 5 & 6 & 7 & 8 & 9 & 10 \\   \hline
   (15) &  11 &   17  &   21 &   25 &   27 &   29 &   30  &  31 &   31 &   32 \\
   \hline
 \end{tabular}
\label{table2}
\end{center}
\end {table}

\section{Conclusion}

In this paper, we formulated the power minimization SINR-constrained problem such that it includes the power dissipated in the RF circuits powering the antennas. The number of active antennas can be modeled using the zero-norm of the antenna's power. We used the standard weighted one-norm approach to replace the zero-norm.  We provided iterative closed-form expressions for our proposed algorithm.  We then extended the algorithm to the case where we have per-antenna power constraints (PAPCs) and derived iterative closed-form expressions as well.  Then we examined the case where many bands are on, and provided analogous closed-form expressions. Our simulations show that we can obtain significant power gains compared to the conventional naive approach that only minimizes the transmit power.

\end{document}